\documentclass[11pt]{article}
\usepackage{amssymb}
\usepackage{graphics}
\usepackage{epsfig}
\usepackage{a4wide}
\begin{document}

\title{ QCD corrections to the R-parity violating processes $p\bar{p}/pp \to e\mu+X$ at hadron colliders  } \vspace{3mm}

\author{{ Wang Shao-Ming$^{2}$,  Han Liang$^{2}$, Ma Wen-Gan$^{1,2}$, Zhang Ren-You$^{2}$, and Jiang Yi$^{2}$}\\
{\small $^{1}$CCAST (World Laboratory), P.O.Box 8730, Beijing, 100080, People's Republic of China} \\
{\small $^{2}$Department of Modern Physics, University of Science and Technology of China (USTC),} \\
{\small       Hefei, Anhui 230026, People's Republic of China} }

\date{}
\maketitle \vskip 12mm

\begin{abstract}
We present the QCD corrections to the processes $p\bar{p}/pp \to
e\mu+X$ at the Tevatron and the CERN large hadron collider(LHC).
The numerical results show that variation of K factor is in the
range between $1.28(1.32)$ and $1.79(1.58)$ at the Tevatron(LHC).
We find that the QCD correction part from the one-loop gluon-gluon
fusion subprocess is remarkable at the LHC and should be taken
into account.
\par
\end{abstract}

\vskip 5cm


\vfill \eject

\baselineskip=0.32in

\renewcommand{\theequation}{\arabic{section}.\arabic{equation}}
\renewcommand{\thesection}{\Roman{section}.}
\newcommand{\nb}{\nonumber}

\makeatletter      
\@addtoreset{equation}{section}
\makeatother       

\par
In the R-parity violating minimal supersymmetric standard model,
the most general representations of superpotential can be written
as
\begin{equation}
\label{sup} {\cal W}_{\rlap/R_{p}} = \frac{1}{2}\epsilon_{ab}
\lambda_{ijk}\hat{L}_{i}^a \hat{L}_{j}^b \hat{E}_{k} +
\epsilon_{ab}\lambda^{'}_{ijk} \hat{L}_{i}^a \hat{Q}_{j}^b
\hat{D}_{k} +
\frac{1}{2}\epsilon_{\alpha\beta\gamma}\lambda^{''}_{ijk}
   \hat{U}_{i}^{\alpha} \hat{D}_{j}^{\beta} \hat{D}_{k}^{\gamma} +
\epsilon_{ab}\delta_{i} \hat{L}_{i}^a \hat{H}_{2}^b
\end{equation}
where $i,j,k$ = 1,2,3 are generation indices; $a,b$ = 1,2 are
SU(2) isospin indices and $\alpha,\beta,\gamma$ are SU(3) color
indices. $\lambda,\lambda^{\prime},\lambda^{\prime\prime}$ are
dimensionless R-violating Yukawa couplings behaving as
$\lambda_{ijk}=-\lambda_{jik}$ and
$\lambda^{\prime\prime}_{ijk}=-\lambda^{\prime\prime}_{jik}$. In
above superpotential, the trilinear terms only violate either L-
or B-symmetry respectively, and the terms which may produce both
L- and B-violation simultaneously, are absent so that a stable
proton is ensured.

\par
An observation of electron-muon pair events with high invariant
mass at hadron colliders would provide evidence of R-parity
violating(RPV) interactions. The electron-muon pair productions at
hadron colliders induced by RPV interactions at the leading order
were investigated in Ref.\cite{ppem}. In this report, we present
the QCD corrections to the RPV  processes $p\bar{p}/pp \to e\mu+X$
at hadron colliders including the contributions of the NLO QCD and
gluon-gluon fusion subprocess. We adopt the dimensional
regularization(DR) method in $D=4-2 \epsilon$ dimensions to
isolate the ultraviolet(UV), infrared(IR) (soft and collinear)
singularities. In order to remove the UV divergences, we employ
the modified minimal subtraction ($\overline{\rm MS}$) scheme to
renormalize and eliminate UV divergences. After renormalization
procedure, the virtual correction part of the cross section is
UV-finite. The IR divergences from the one-loop diagrams involving
gluon will be cancelled by adding the soft real gluon emission
corrections by using the two cutoff phase space slicing method
(TCPSS)\cite{twocut}. The remaining collinear divergences can be
absorbed into the parton distribution functions.

\par
Although the contribution of the $e\mu$ production via gluon-gluon
fusion at the lowest order is an one-loop subprocess and this
contribution part to the process $p\bar {p}(pp) \to e\mu+X$ is at
${\cal O}(\alpha_s^2)$ order, which is higher than that of the NLO
QCD correction to the process $p\bar p(pp) \to q \bar q \to
e\mu+X$, it is possible that the production rate of the $p\bar
p(pp) \to gg \to e\mu+X$ could be non-negligible in contrast with
the NLO QCD correction to the process via $q\bar q$ annihilation ,
due to large gluon luminosity in TeV-scale
proton-proton(anti-proton) collisions. In this report we present
also the contribution part via gluon-gluon fusions.

\par
In the numerical calculations of the QCD corrected cross sections
at the Tevatron and the LHC, we take the RPV parameters $\lambda$
and $\lambda^{'}$ to be real for simplicity with the values as
\begin{eqnarray}
\lambda_{112}=\lambda_{221}=0,~~
\lambda_{212}=\lambda_{121}=0.049,~~\lambda_{312}=0.062,~~\lambda_{321}=0.070, \nb \\
\lambda'_{111}=5.2\times10^{-4},~~\lambda'_{112}=\lambda'_{113}=0.021, \nb \\
\lambda'_{121}=0.043,~~\lambda'_{131}=0.019,~~\lambda'_{211}=\lambda'_{212}=\lambda'_{213}=0.059,
\nb \\
\lambda'_{221}=\lambda'_{231}=0.18,~~\lambda'_{311}=0.11,
\end{eqnarray}
where the values of $\lambda$ and $\lambda^{'}$ are under the
experimental constraints presented in Ref.\cite{sneu}. We use the
CTEQ6L parton distribution function for the tree-level cross
sections and CTEQ6M for one-loop QCD corrected cross section
results \cite{pdfs}. The factorization and the renormalization
scales are set to be equal and taken as $\mu_f = \mu_r =
m_{\tilde\nu}$. We applied the naive fixed-width scheme and fix
the decay width of the sneutrino propagator being $\Gamma=10~GeV$,
to avoid the possible resonant singularities. Since the sneutrinos
are non-colored supersymmetric particles, there is no problem with
gauge invariance or double counting of higher-order effects in
calculating the cross sections of $q \bar q \to e\mu$ involving
the NLO QCD corrections. The gluino and squark masses are set to
be $m_{\tilde g}=916.1GeV$ and $m_{\tilde{q}}=200(900)~GeV$, and
$2 \times 2$ mixing matrices $R^{\tilde u}$ and $R^{\tilde d}$ are
taken to be unit for simplification. By using the TCPSS method, we
take the soft cutoff $\delta_s=10^{-2}$ and collinear cutoff
$\delta_c=\delta_s/50$. The calculations are carried out at the
Tevatron with $p\bar{p}$ colliding energy $\sqrt{s} = 1.96~TeV$,
and at the LHC with $pp$ colliding energy $\sqrt{s} = 14~TeV$.
Since the $\overline{MS}$ scheme violates supersymmetry, the
$q\tilde{q}\tilde{g}$ Yukawa coupling $\hat{g}_s$ takes a finite
shift at one-loop order as shown in Eq.(\ref{shift})
\cite{shiftgs}:
\begin{eqnarray}
\label{shift} && \hat{g}_s = g_s \left [1
+\frac{\alpha_s}{8\pi}\left (\frac{4}{3}N_c - C_F\right )\right ],
\end{eqnarray}
with $N_c=3$ and $C_F=4/3$. In our numerical calculation we take
this coupling strength shift between $\hat{g}_s$ and $g_s$ into
account.

\par
In Fig.1(a) and (b) we depict the curves of the tree-level and QCD
corrected cross sections($\sigma^{0}$ and $\sigma^{QCD}$) of the
processes $p\bar{p}/pp \to e^+\mu^-+X$ involving NLO QCD and
gluon-gluon fusion subprocess corrections versus the sneutrino
mass $m_{\tilde{\nu}}$ with squark mass being $200~GeV$ and
$900~GeV$ at the Tevatron and the LHC, respectively. Their
corresponding K factors($K \equiv
\frac{\sigma^{QCD}}{\sigma^{0}}$) as the functions of
$m_{\tilde{\nu}}$ are depicted in Fig.1(c) and (d), separately. We
can see the cross section curves in Fig.1(a,b) decrease rapidly
with the increment of $m_{\tilde{\nu}}$, and the QCD corrected
cross sections for a sneutrino with several hundred GeV mass are
changed slightly when the $m_{\tilde{q}}$ value runs from
$200~GeV$ to $900~GeV$, due to the decrease of contribution from
the squark exchanging ones. We can read out from Fig.1(c-d) that
the K factors vary in the ranges of $[1.28,1.79]$ at the Tevatron
and $[1.32,1.58]$ at the LHC. For a $100~GeV$ sneutrino and
$900~GeV$ squarks, the K factors reach 1.79 and 1.58 at the
Tevatron and the LHC, respectively.

\begin{figure}[htp]
\includegraphics[width=3.2in,height=3in]{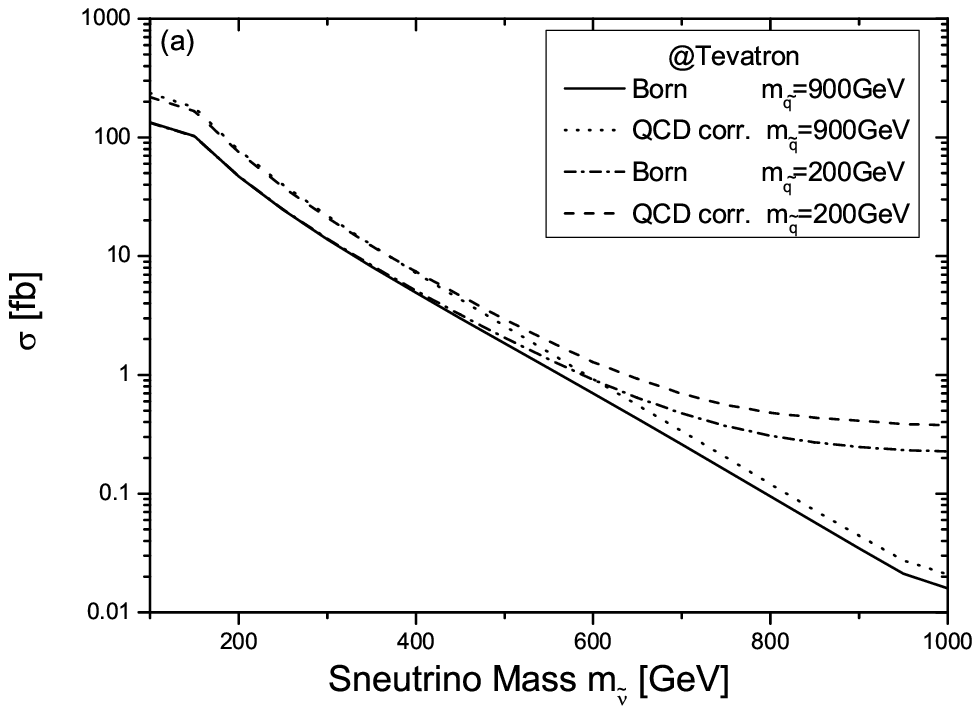}%
\hspace{0in}%
\includegraphics[width=3.2in,height=3in]{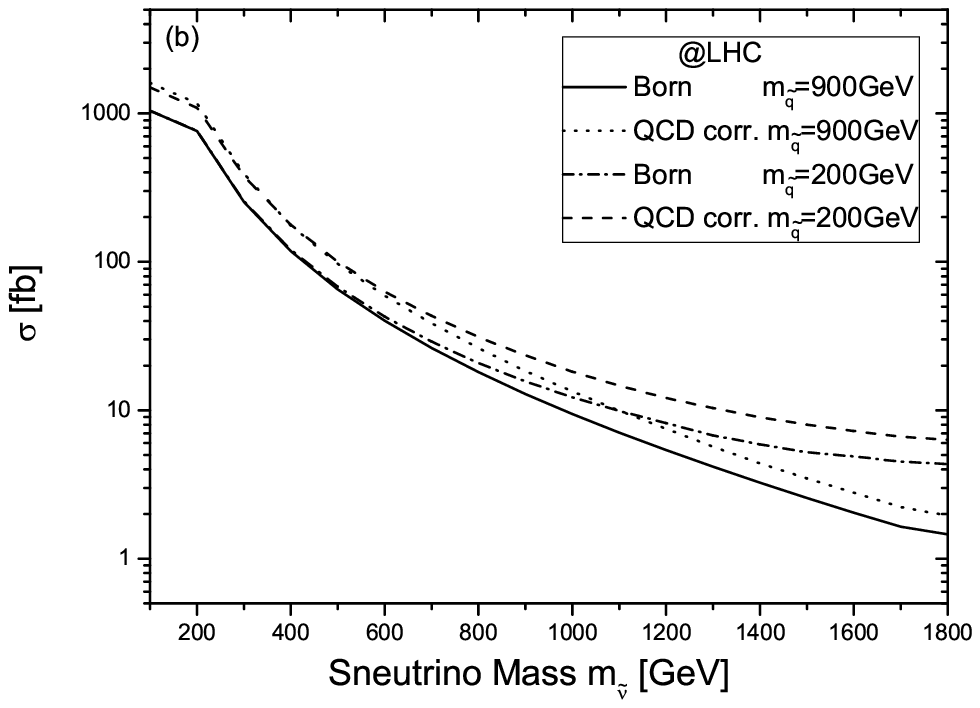}%
\hspace{0in}%
\includegraphics[width=3.2in,height=3in]{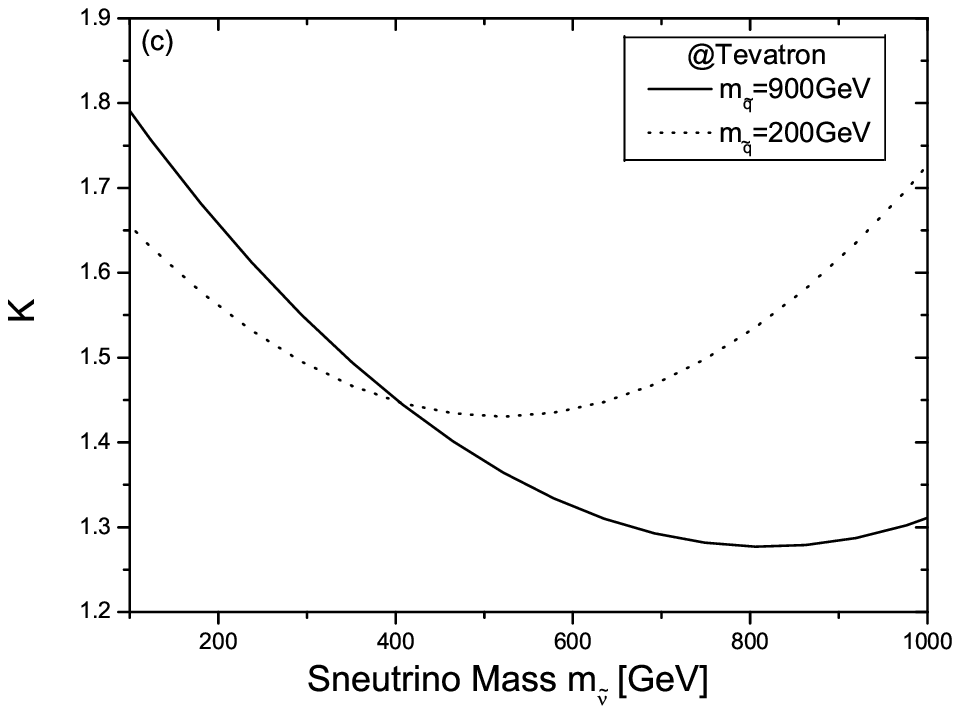}%
\hspace{0in}%
\includegraphics[width=3.2in,height=3in]{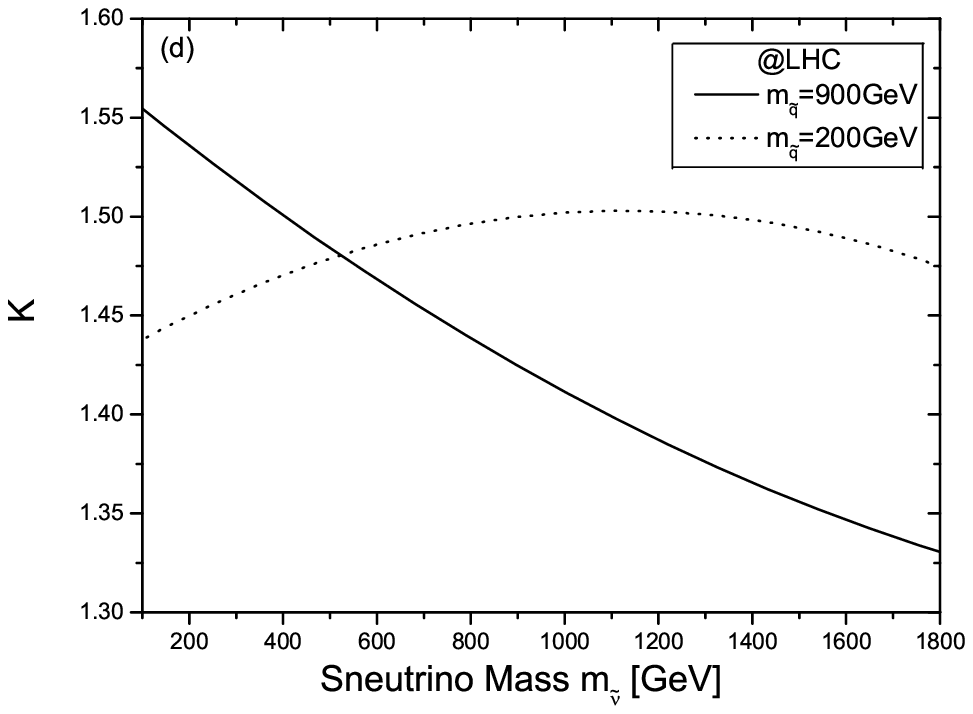}%
\caption{The tree-level and total QCD corrected cross sections
($\sigma^{0}$ and $\sigma^{QCD}$) of the processes $p\bar{p}/pp
\to e^+\mu^-+X$ involving NLO QCD and gluon-gluon fusion
subprocess corrections at the Tevatron and the LHC as the
functions of the sneutrino mass $m_{\tilde{\nu}}$ with different
squark masses are shown in Fig.7(a) and (b), respectively.
Fig.7(c-d) show the corresponding relations between the K factors
of the processes and the sneutrino mass $m_{\tilde{\nu}}$. (c) at
the Tevatron, (d) at the LHC. }
\end{figure}

\par
In Table 1 we list the tree-level cross sections of the processes
$p\bar p/pp \to e^+\mu^-+X$, their K factors and the contributing
parts from gluon-gluon fusion mechanism in conditions of taking
different mass values of sneutrino and squark at the Tevatron and
the LHC, with positron transverse momentum $p_T >
p_T^{cut}=20~GeV$ at the Tevatron and $p_T > p_T^{cut}=25~GeV$ at
the LHC. We can read from Table 1 that the relative corrections
from the gluon-gluon fusion subprocess at the Tavatron is less
than $1\%$ and can be negligible, but at the LHC the relative
corrections from the gluon-gluon fusion subprocess can reach $6\%$
when $m_{\tilde{\nu}}$ has the value of $200~GeV$ due to large
gluon luminosity at the LHC. We can see that in evaluating the QCD
corrections to the processes $p\bar p/pp \to e^+\mu^-+X$ at the
LHC, it is reasonable to take the contribution from the
gluon-gluon fusion subprocess into account.

\vskip 50mm
\begin{table}[htb]
\centering
\begin{tabular}{|c|c|c|c|c|c|c|c|c|}
\hline $m_{\tilde{\nu}}$ & $m_{\tilde{q}}$ &
$\sigma^{0}_{Tevatron}$ & $K_{Tevatron}$ &
$\sigma^{gg}_{Tevatron}/\sigma^{0}_{Tevatron}$ &
$\sigma^{0}_{LHC}$ &
$K_{LHC}$ & $\sigma^{gg}_{LHC}/\sigma^{0}_{LHC}$\\
\hline
200  & 200 & 39.83 & 1.5737 & 0.00487 & 467.5 & 1.525 & 0.0598 \\
\hline
500  & 200 & 1.925 & 1.4312 & 0.000741 & 61.25 & 1.477 & 0.0107 \\
\hline
900  & 200 & 0.2436 & 1.6671 & 0.00229 & 15.13 & 1.520 & 0.0144 \\
\hline
200  & 900 & 39.60 & 1.6421 & 0.00488 & 464.9 & 1.630 & 0.0595 \\
\hline
500  & 900 & 1.708 & 1.3954 & 0.000461 & 58.60 & 1.489 & 0.00707 \\
\hline
900  & 900 & 0.03171 & 1.2778 & 0.00195 & 12.37 & 1.433 & 0.00154 \\
\hline
\end{tabular}
\caption{\small The tree-level cross sections of the processes
$p\bar{p}/pp \to e^+\mu^-+X$(in fb), the K factor and the relative
correction from the gluon-gluon fusion subprocess with different
mass values(GeV) of the sneutrino and the squark at the Tevatron
and the LHC, in conditions of positron transverse momentum $p_T >
p_T^{cut}=20~GeV$ for the Tevatron and $p_T
> p_T^{cut}=25~GeV$ for the LHC.}
\end{table}

\par
We present the plots of the differential cross
sections($d\sigma^{0}/dp_T$ and $d\sigma^{QCD}/dp_T$) for the
processes $p\bar{p}/pp \to e^+\mu^-+X$ versus the transverse
momentum $p_T$ of outgoing positron in Fig.2(a) and (b) at the
Tevatron and the LHC respectively in conditions of
$m_{\tilde{\nu}} = 250~GeV$ and $m_{\tilde{q}} = 200~GeV$. We find
that in the high $p_T$ region the relative QCD correction to the
the differential cross sections can be more significant,
especially at the LHC.

\begin{figure}[htbp]
\includegraphics[width=3.2in,height=3in]{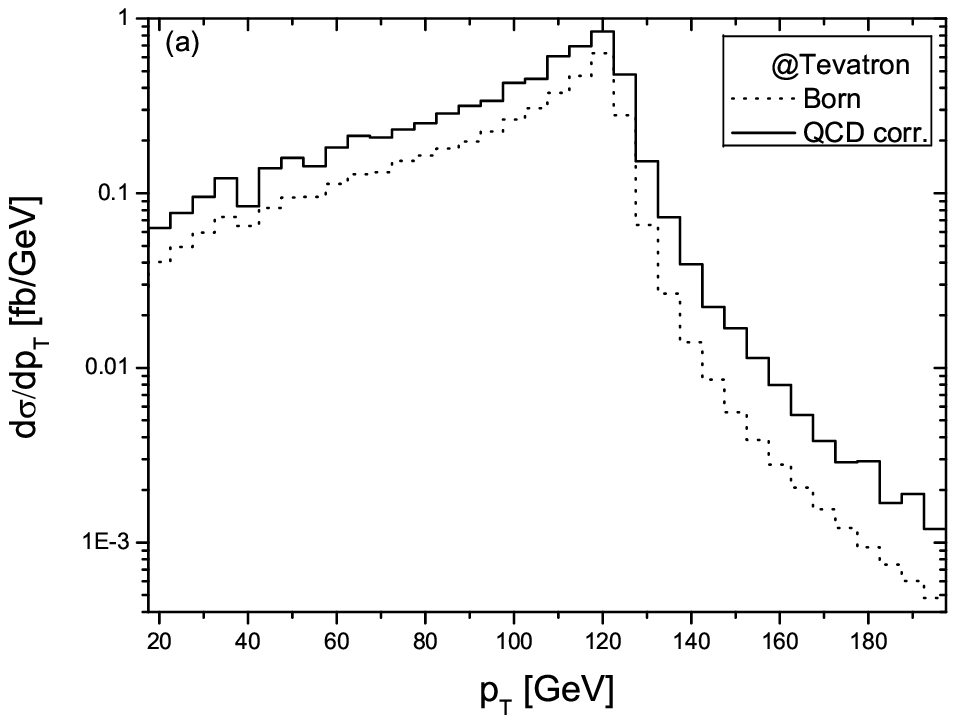}%
\hspace{0in}%
\includegraphics[width=3.2in,height=3in]{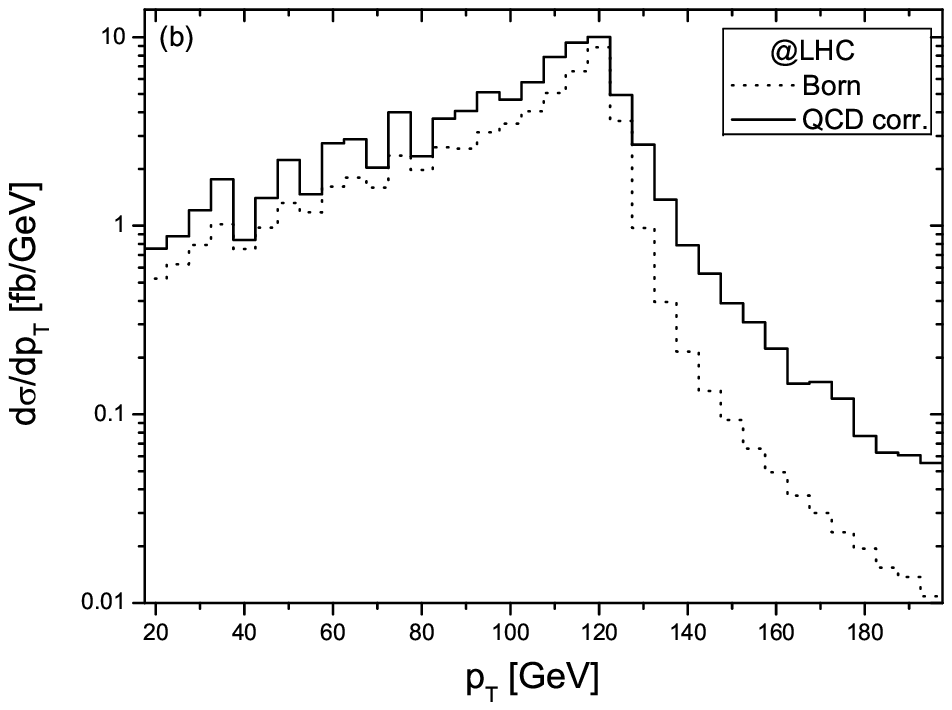}%
\hspace{0in}%
\caption{The distributions of the transverse momentum of outgoing
positron $d\sigma^{0}/dp_T$ and $d\sigma^{QCD}/dp_T$ for the
processes $p\bar{p}/pp \to e^+\mu^-+X$. (a) at the Tevatron, (b)
at the LHC. }
\end{figure}

\par
In summary, we studied the QCD corrections to the lepton flavor
violating processes $p\bar{p}/pp \to e\mu+X$ at the Tevatron and
the LHC including the one-loop QCD corrections to $q \bar q \to
e\mu$ subprocess and the one-loop subprocess $gg \to e\mu$. In our
investigating parameter space the K factors vary in the ranges of
$[1.28,1.79]$ at the Tevatron and $[1.32,1.58]$ at the LHC. For a
$100~GeV$ sneutrino and $900~GeV$ squarks, the K factors reach
$1.79$ and $1.58$ at the Tevatron and the LHC, respectively. We
find that the contribution to the total QCD correction from the
one-loop gluon-gluon fusion subprocess is remarkable at the LHC,
and its relative correction can reach $6\%$ at the LHC when
$m_{\tilde{\nu}}=m_{\tilde{q}}=200~GeV$.

\vskip 5mm
\par
\noindent{\large\bf Acknowledgments:} This work was supported in
part by the National Natural Science Foundation of China, the
Education Ministry of China and a special fund sponsored by
Chinese Academy of Sciences.


\vskip 10mm
\begin{thebibliography}{s25}
\bibitem{ppem} Sun Yan-Bin, {\textit et al}., Commun. Theor. Phys. {\bf 44}, 107-116(2005), hep-ph/0412205.
\bibitem{twocut}B. W. Harris and J.F. Owens, Phys. Rev. {\bf D65} (2002) 094032, hep-ph/0102128.
\bibitem{sneu} R. Barbieri, {\textit et al}., hep-ph/9810232; B. Allanach et al., hep-ph/9906224;
    F. Deliot, {\textit et al}., Phys. Lett. {\bf B475} (2000)184; G. Moreau, {\textit et al}., Nucl. Phys. {\bf B604} (2001)3;
    S. Bar-Shalom, G. Eilam and B. Mele, Phys. Rev. {\bf D64} (2001) 095008.
\bibitem{pdfs} J. Pumplin, \textit{et al}., JHEP 0207, 012 (2002); D. Stump, \textit{et al}., JHEP 0310, 046 (2003).
\bibitem{shiftgs} W. Beenakker, R. H$\ddot{\rm o}$pker, P.M. Zerwas, Phys. Lett. {\bf B378} (1996) 159;
         W. Beenakker, R. H$\ddot{\rm o}$pker, T. Plehn, P.M. Zerwas, Z. Phys. {\bf C75} (1997) 349.
\end{thebibliography}
\end{document}